\documentclass{article}

\usepackage[main, final]{neurips_2026}

\usepackage[utf8]{inputenc}
\usepackage[T1]{fontenc}
\usepackage{float}
\usepackage[hidelinks]{hyperref}
\usepackage{url}
\usepackage{booktabs}
\usepackage{amsfonts}
\usepackage{amsmath}
\usepackage{amssymb}
\usepackage{nicefrac}
\usepackage{microtype}
\usepackage{xcolor}
\usepackage{graphicx}
\usepackage{afterpage}
\usepackage{algorithm}
\usepackage{algpseudocode}
\usepackage{bm}

\title{GlycoGen: Glycan Structure Prediction \\
       with Crystallizing Flows}

\author{%
  Romain Lacombe \\
  Independent Researcher \\
  \texttt{romain@romainlacombe.com}
}

\begin{document}

\maketitle

\begin{abstract}
Glycosylation is among the most abundant post-translational
modifications in biology, governing cellular signaling and the
foundational mechanisms of immunity. Because glycans cannot be
sequenced from a genetic template, their structure must be inferred
directly from tandem mass spectrometry (MS/MS).
Modern instruments acquire 100 to 200 spectra per second, yet the
prior open-vocabulary state of the art, GlycoBART, is a 207M
autoregressive transformer whose sequential decoding precludes live
acquisition and large-scale glycomics.
Discrete Flow Matching and masked diffusion remove this sequential
bottleneck by generating all token positions in parallel, provided a
low-step sampler preserves structural accuracy.
We introduce \textbf{GlycoGen}, a 50M-parameter Discrete Flow Matching
model over fragment spectra, paired with \textbf{Crystallizing Flows},
a deterministic sampler that runs out-of-the-box via mixture
embeddings with no retraining. Crystallizing Flows integrate per-token
distributions on the simplex, snap each position to a one-hot vector
once its argmax leaves the mask, and exit early once all positions
crystallize.
On the 76,774-spectrum CandyCrunch~v2 benchmark, GlycoGen reaches
\textbf{92.67\%} top-1 accuracy in 17 passes and \textbf{90.19\%} in a
single pass, surpassing GlycoBART by 11.7 percentage points with a
14$\times$ to 61$\times$ speedup.
Setting a new state of the art in open-vocabulary glycan structure
prediction, GlycoGen makes the glycome readable in real time and at
scale.
\end{abstract}

\section{Introduction}\label{sec:intro}

\begin{figure}[t]
\centering
\includegraphics[width=0.95\textwidth]{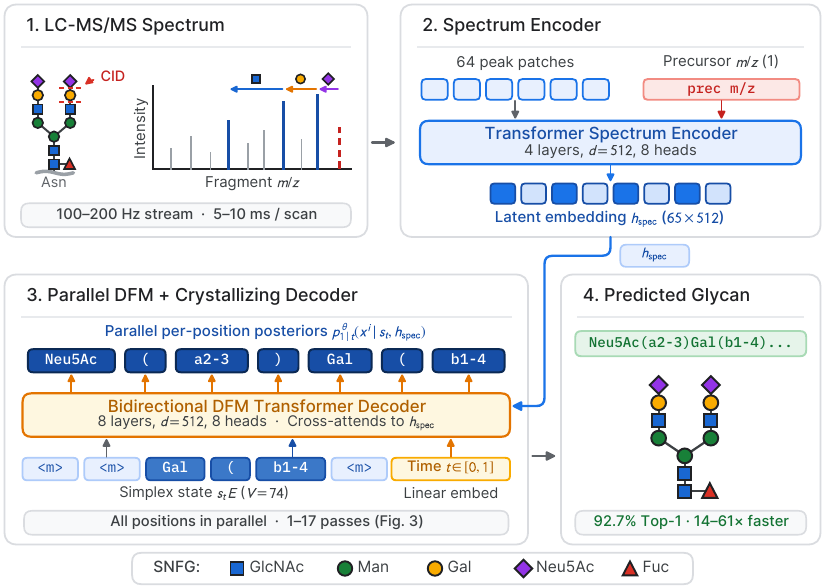}
\caption{\textbf{GlycoGen end-to-end pipeline.}
\textbf{(1)} Low-energy CID cleaves glycosidic bonds of branched glycans
(SNFG notation~\citep{varki2015snfg}) into diagnostic monosaccharide
losses ($\Delta m/z$) streamed at 100--200\,Hz.
\textbf{(2)} A 4-layer Transformer Spectrum Encoder maps 64 binned
intensity/mass-offset patches and one precursor $m/z$ token into
$\mathbf{h}_\mathrm{spec} \in \mathbb{R}^{65 \times 512}$.
\textbf{(3)} An 8-layer bidirectional DFM decoder maps simplex mixture
embeddings $\mathbf{s}_t E$ and time $t \in [0, 1]$ to all 74-token
atomic IUPAC positions in parallel.
\textbf{(4)} Crystallizing Flows (Figure~\ref{fig:method}) reach 92.7\%
accuracy at a 14$\times$--61$\times$ speedup.}
\label{fig:pipeline}
\end{figure}

Nearly all cell-surface and secreted proteins are glycosylated,
carrying branched sugar structures that govern cell signaling,
self--nonself recognition, and immune regulation.
Unlike proteins and nucleic acids, glycans are assembled by competing
enzymes without a genetic template; a single monosaccharide composition
can form dozens of linkage isomers with identical mass. Tandem mass
spectrometry (MS/MS) is the primary
experimental probe for resolving these structures. Mapping a
fragment-ion spectrum back to its underlying glycan tree is a
combinatorial inverse problem, making structure prediction a central
computational bottleneck in glycomics---the large-scale, systematic
study of glycans (Figure~\ref{fig:pipeline}).

This bottleneck spans both live instrument control and offline
analysis. Modern MS/MS instruments acquire 100 to 200 spectra per
second; data-dependent acquisition can steer next-scan selection or
collision energy on the fly only when inference finishes within the 5
to 10 millisecond scan interval, as practiced via fast search in
proteomics~\citep{schweppe2020rts, furtwangler2022rts} and
glycoproteomics~\citep{armony2023glycopaser, sutherland2024rtls}. Even offline, a
biomedical characterization campaign generates millions of spectra;
slow structure prediction bottlenecks analysis and routinely leaves
rare or low-abundance spectra unannotated. Spectral library search is
fast only for catalogued structures, while the prior neural state of
the art, GlycoBART~\citep{glycobart2025}, decodes compressed antenna
motifs autoregressively one token per forward pass---scaling latency
linearly with sequence length, losing structural accuracy when
reconstructing branched trees, and relying on database snap-back over
known glycans. Open-vocabulary glycan MS/MS thus faces a dual
bottleneck of sequential latency and lossy structure generation.

To address both challenges, we introduce \textbf{GlycoGen}, a
50M-parameter Discrete Flow Matching denoiser~\citep{gat2024dfm} that
predicts open-vocabulary glycan structures directly from full fragment
spectra without database lookup. Conditioning on binned peak
intensities and sub-bin mass offsets across the entire spectrum,
GlycoGen generates atomic IUPAC tokens at all sequence positions
simultaneously, decoupling inference latency from glycan length while
letting every position attend bidirectionally across sibling branches.

Parallel generation realizes its speed advantage only if a low-step
sampler preserves structural accuracy, yet standard Continuous-Time
Markov Chain (CTMC) Euler sampling~\citep{gat2024dfm} draws random
categorical commits that degrade sharply at small step budgets. We
therefore pair GlycoGen with \textbf{Crystallizing Flows}
(Figure~\ref{fig:method}), a parameter-free deterministic sampler that
integrates per-position distributions on the probability simplex via
mixture embeddings, snaps each position to a one-hot token once its
argmax leaves \texttt{<mask>}, and exits early once all positions
crystallize. \textbf{Together, GlycoGen and Crystallizing Flows set a
new state of the art in open-vocabulary glycan structure prediction},
surpassing both our matched autoregressive control and GlycoBART in
accuracy and speed (Table~\ref{tab:results}, Figure~\ref{fig:pareto-ms}).

\paragraph{Contributions.}

\noindent\textbf{(i) GlycoGen: state-of-the-art open-vocabulary glycan
structure prediction.} On the 76,774 test spectra of the
CandyCrunch~v2 benchmark~\citep{candycrunch2024}, single-pass GlycoGen
reaches 90.19\% top-1 accuracy across three seeds, surpassing greedy
GlycoBART (80.92\%) by 9.3 percentage points at $61\times$ batched
throughput and $21\times$ lower serial latency.

\noindent\textbf{(ii) Crystallizing Flows: training-free deterministic
simplex sampling.} This parameter-free sampler reaches
\textbf{92.67\%} top-1 accuracy in 17 effective passes (92.51\% in 5
passes), Pareto-dominating standard CTMC Euler at low step counts and
transferring without retraining as a drop-in replacement for
stochastic sampling on a Masked Diffusion Language Model (MDLM)
denoiser~\citep{sahoo2024mdlm}.

\noindent\textbf{(iii) Graph-isomorphic generalization and
single-spectrum Bayes error ceiling.}
On test glycans that appear in training only under a different IUPAC
branch ordering, GlycoGen reaches 60.6\% accuracy (62.1\% with MDLM)
versus 13.6\% for GlycoBART. On seen glycan structures, GlycoGen
reaches 93.31\% graph accuracy---approaching the Bayes error ceiling of
single-spectrum CID, as four in five remaining errors reflect linkage
isomers or labile sugar losses that single-stage MS/MS cannot physically
distinguish (Section~\ref{sec:results}).

\section{Background and Prior Work}\label{sec:background}

\subsection{Glycomics and Tandem Mass Spectrometry}\label{sec:glyco-bg}

\begin{figure}[t]
\centering
\includegraphics[width=0.94\textwidth]{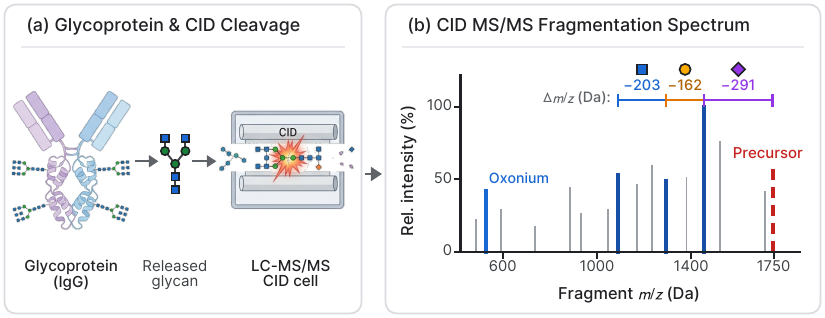}
\caption{\textbf{Biological primer on glycomics LC-MS/MS.} Branched
N-glycans released from glycoproteins (e.g., IgG antibodies) enter the
collision cell, where low-energy CID cleaves glycosidic bonds into
fragment spectra plotting relative intensity against $m/z$. Sequential
mass intervals ($\Delta m/z$) read off monosaccharide losses
($-291$\,Da Neu5Ac, $-162$\,Da Gal, $-203$\,Da GlcNAc) from which
GlycoGen reconstructs the glycan.}
\label{fig:biorender-primer}
\end{figure}

\paragraph{Glycan biology and CID fragmentation.}
Glycans are central to fundamental functions across biology, from
cell--cell recognition and receptor signaling to viral entry, mucosal
barrier defense, and the fine-tuning of antibody effector function.
Assembled without a genetic template by competing glycosyltransferases
and glycosidases, two classes dominate mammalian glycoproteins:
asparagine-linked N-glycans built around a conserved
chitobiose-trimannosyl core and serine/threonine-linked O-glycans built
on mucin cores. In liquid chromatography--tandem mass spectrometry
(LC-MS/MS), glycans separated by retention time are isolated at a
precursor $m/z$ and fragmented by low-energy collision-induced
dissociation (CID). In the nomenclature of \citet{domon1988}, low-energy
CID predominantly cleaves glycosidic bonds ($B$- and $Y$-ions) whose
$\Delta m/z$ intervals identify lost monosaccharides
(Figures~\ref{fig:pipeline} and~\ref{fig:biorender-primer}), whereas
higher-energy cross-ring cleavages ($A$- and $X$-ions) needed to separate
regioisomers are weak or absent, limiting single-scan isomer resolution.

\paragraph{Glycan representations.}
Glycan structures are depicted in Symbol Nomenclature for Glycans
(SNFG;~\citealp{varki2015snfg, neelamegham2019snfg}) and serialized as
condensed IUPAC strings where parentheses specify glycosidic linkages
and brackets enclose branches (Figure~\ref{fig:pipeline}). Because
sibling branches lack a unique linear order, distinct strings can encode
the same molecular graph (Figure~\ref{fig:qual-snfg}a;
Appendix~\ref{app:inst-gen}), motivating graph-isomorphism scoring in
\texttt{glycowork}~\citep{glycowork2021}. Public repositories catalog
these structures and spectra across
GlyTouCan~\citep{tiemeyer2017glytoucan, fujita2021glytoucan},
GlyConnect~\citep{alocci2019glyconnect},
GlycoPOST~\citep{watanabe2021glycopost}, and the 500,000-spectrum
CandyCrunch corpus~\citep{candycrunch2024, zenodo2024cc2}.

\subsection{Prior Work}\label{sec:prior-work}

\paragraph{Discrete Flow Matching and masked diffusion.}
Continuous diffusion~\citep{ho2020ddpm} and Flow
Matching~\citep{lipman2023fm} learn velocity fields transporting a
simple source to the data distribution. \citet{campbell2022ctmc,
campbell2024discrete} formulated continuous-time categorical flows via
CTMCs, and \citet{gat2024dfm} generalized them into Discrete Flow
Matching across arbitrary sources, couplings, and schedules. Related
absorbing-state masked diffusion models---D3PM~\citep{austin2021d3pm},
SEDD~\citep{lou2024sedd}, MDLM~\citep{sahoo2024mdlm}, and
MD4~\citep{shi2024md4}---match autoregressive language models at
scale~\citep{nie2025llada} and outperform them under data
scarcity~\citep{prabhudesai2025diffusion}. Unlike simplex-native
training relaxations~\citep{dieleman2022cdcd, davis2024fisherfm,
stark2024dirichlet}, Crystallizing Flows integrate on the simplex purely
at inference time atop discretely trained checkpoints
(Appendix~\ref{app:samplers}).

\paragraph{Inference-time acceleration.}
Continuous diffusion acceleration uses higher-order ODE
solvers~\citep{song2020ddim, lu2022dpmsolver, karras2022edm},
distillation~\citep{salimans2022progressive, song2023consistency,
luo2023lcm, meng2023guided, sauer2023add, yin2023dmd,
lacombe2024distillation}, and rectified flows~\citep{liu2022rectified}.
Discrete unmasking has relied on heuristic confidence schedules such as
Mask-Predict~\citep{ghazvininejad2019maskpredict} and
MaskGIT~\citep{chang2022maskgit} or stochastic CTMC
stepping~\citep{sahoo2024mdlm, gat2024dfm}; Crystallizing Flows instead
pair simplex propagation with a parameter-free argmax commit rule.

\paragraph{Computational mass spectrometry.}
Classical glycan tools match fragments against rules or curated
libraries, including GlycoWorkbench~\citep{ceroni2008glycoworkbench},
GlycoDeNovo~\citep{hong2017glycodenovo}, pGlyco3~\citep{pglyco2021}, and
StrucGP~\citep{shen2021strucgp}. Neural decoders such as
Casanovo~\citep{casanovo2022, yilmaz2024casanovo},
InstaNovo~\citep{instanovo2025}, and DiffMS~\citep{diffms2025} predict
peptides and small molecules directly from MS/MS spectra. In glycomics,
CandyCrunch~\citep{candycrunch2024} classifies over a fixed 3,500-glycan
library, leaving 207M autoregressive GlycoBART~\citep{glycobart2025,
lewis2020bart} as the sole open-vocabulary neural baseline. Compact
IUPAC vocabularies, short branched trees (10--50 tokens), and
millisecond scan budgets make glycan MS/MS well suited to parallel flow
matching.

\section{GlycoGen and Crystallizing Flows}\label{sec:method}

\subsection{Mask-Source Discrete Flow Matching}\label{sec:dfm}

Discrete Flow Matching~\citep{gat2024dfm} defines a probability path
$\{p_t\}_{t \in [0,1]}$ between a tractable source $p_0$ and the data
distribution $p_1 = q$, then trains a denoiser
$p_{1 \mid t}^{\theta}$ whose generating velocity transports samples
along that path. We use the mask source $p(x) = \delta_m(x)$ with
the linear schedule $\kappa_t = t$, which yields the per-position
mixture path
\begin{equation}
  p_t(x^i \mid x_0, x_1) =
    (1 - t)\, \delta_m(x^i) + t\, \delta_{x_1^i}(x^i),
  \label{eq:dfm-path}
\end{equation}
the masked cross-entropy training loss
\begin{equation}
  \mathcal{L}(\theta) \;=\;
    -\, \mathbb{E}_{t,\, X_1,\, X_t}\!
    \sum_{i \in \mathcal{M}_t}
    \log p_{1 \mid t}^{\theta}(X_1^i \mid X_t, \mathbf{h}_\mathrm{spec}),
  \quad t \sim \mathcal{U}(0,1),
  \label{eq:dfm-loss}
\end{equation}
and the closed-form per-position velocity~\citep{gat2024dfm}
\begin{equation}
  u_t^i(x, z) \;=\;
    \tfrac{1}{1 - t}\bigl[\, p_{1 \mid t}^{\theta}(x \mid z, \mathbf{h}_\mathrm{spec})
    - \delta_{z^i}(x)\, \bigr].
  \label{eq:dfm-velocity}
\end{equation}
The standard CTMC Euler sampler of \citet{gat2024dfm}, reproduced in
Appendix~\ref{app:alg1}, integrates this velocity stochastically over
a uniform grid of $K$ steps with step size $h = 1/K$. At each step,
the state is a discrete token sequence, and each masked position
commits to a sampled token with probability $h/(1-t_k)$ and stays
frozen thereafter. Because every commit collapses the softmax
distribution to a categorical draw and cannot be revised, early
errors condition later steps.

Crucially, the training objective in Equation~\eqref{eq:dfm-loss}
leaves the inference sampler unspecified: any procedure that integrates
the velocity field from the all-mask source at $t = 0$ to a clean
sequence at $t = 1$ is a valid decoder. This decoupling between
training and inference has motivated dedicated research into discrete
unmasking schedules and samplers~\citep{ghazvininejad2019maskpredict,
chang2022maskgit, sahoo2024mdlm, gat2024dfm}; we build on it directly
in Section~\ref{sec:adaptive}.

\subsection{GlycoGen Architecture}\label{sec:arch}

Two architectural choices separate GlycoGen from prior glycan structure
predictors. First, GlycoGen is \textbf{open-vocabulary}: rather than
classifying over a fixed glycan dictionary~\citep{candycrunch2024} or
predicting compressed antenna motifs whose reconstruction loses
branch-attachment accuracy and relies on database
snap-back~\citep{glycobart2025}, GlycoGen decodes directly over a
74-token atomic IUPAC vocabulary (70 monosaccharide, linkage such as
\texttt{a1-3} or \texttt{b1-4}, and branch-bracket \texttt{[}/\texttt{]}
tokens, plus \texttt{<pad>}, \texttt{<mask>}, \texttt{<bos>}, and
\texttt{<eos>}). Any chemically valid branched glycan tree in this
alphabet can be generated directly. Second, GlycoGen is
\textbf{non-autoregressive}: whereas causal decoders require $L \approx
33$ sequential forward passes per beam step ($33 \times B$ passes at
beam width $B$), GlycoGen applies full bidirectional self-attention
across all $L = 96$ sequence slots simultaneously, refining the entire
branched tree in parallel in $K \ll L$ passes (down to $K = 1$).

Each MS/MS spectrum records a variable-length list of fragment-ion
$(m/z, \text{intensity})$ peaks produced when the instrument fragments
an isolated precursor ion, where low-intensity diagnostic ions often
resolve branching or linkage isomers. Rather than truncating to a
top-$N$ peak list, we map each spectrum onto a fixed 2,048-bin
mass-to-charge grid storing two normalized channels per bin---log-scaled
peak intensity and continuous sub-bin mass offset---preserving full
spectral coverage without variable-length padding while retaining
sub-bin mass precision (Figure~\ref{fig:spectrum-patching},
Appendix~\ref{app:patching}). The spectrum encoder reshapes these 2,048
dual-channel bins into 64 contiguous patches of 32 bins ($64$ features
per patch), linearly projects each patch to $d = 512$, appends one
standardized precursor-$m/z$ token projected to $d = 512$, adds learned
positional embeddings, and passes the 65-token sequence through a
4-layer pre-norm GELU Transformer encoder~\citep{vaswani2017attention}
(8 attention heads, feed-forward width $2{,}048$) to compute
$\mathbf{h}_\mathrm{spec} \in \mathbb{R}^{65 \times 512}$ once per
spectrum.

The structure decoder embeds the partially masked glycan sequence of
length $L = 96$ via token embedding matrix $E \in \mathbb{R}^{74 \times
512}$ and learned positional embeddings, adds a single linear broadcast
embedding of continuous time $t \in [0, 1]$ across all positions
(avoiding the adaptive layer-norm overhead of
DiT;~\citealp{peebles2023dit}), cross-attends to
$\mathbf{h}_\mathrm{spec}$ across an 8-layer bidirectional pre-norm GELU
Transformer decoder ($d = 512$, 8 heads, feed-forward width $2{,}048$),
and outputs per-position logits over all 74 tokens in parallel.

\subsection{Crystallizing Flows Sampler}\label{sec:adaptive}

\begin{figure}[t]
\centering
\includegraphics[width=0.95\textwidth]{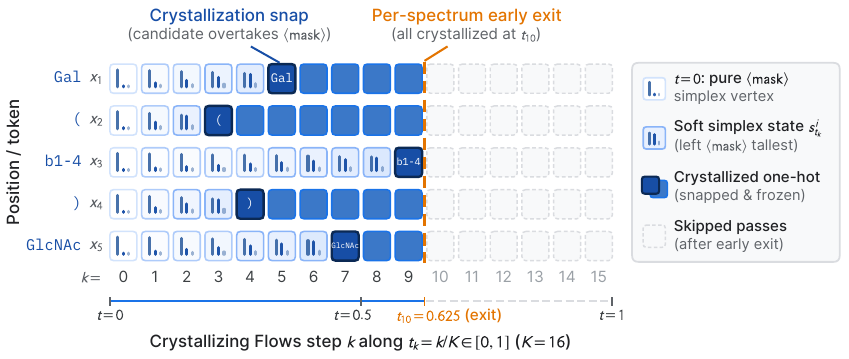}
\caption{\textbf{Crystallizing Flows mechanic.} Per-position state
evolution across 16 Crystallizing Flows steps on \texttt{Gal(b1-4)GlcNAc}.
\textcolor[HTML]{1A73E8}{\textbf{Light-blue cells}} show soft simplex
distributions $\mathbf{s}_{t_k}^i \in \Delta^{V-1}$ whose tallest left
bar remains \texttt{<mask>} while the candidate bar rises prior to
commit. Once the candidate overtakes \texttt{<mask>}, the position snaps
to a \textcolor[HTML]{174EA6}{\textbf{deep-blue one-hot}} token and
stays frozen thereafter. Decoding exits early at the
\textcolor[HTML]{E37400}{\textbf{orange dashed line}} ($t_{10} =
0.625$) once all positions crystallize.}
\label{fig:method}
\end{figure}

At inference time, what state we pass from one Euler step to the next
determines how much of the denoiser's predictive distribution survives
the rollout. Whereas CTMC Euler discards each step's vocabulary
distribution in favor of a random categorical draw and commits masked
positions by coin flip (Section~\ref{sec:dfm}), we \textbf{propagate
each position's full probability distribution on the simplex} and let
the model itself decide when to commit. We call this deterministic
sampler \textbf{Crystallizing Flows} (Figure~\ref{fig:method}): each
undecided position maintains a probability vector over the vocabulary
and snaps to a one-hot token the moment its argmax leaves
\texttt{<mask>}, preserving calibrated uncertainty early in the
trajectory while eliminating low-step sampling variance.

\paragraph{Deterministic integration on the simplex.}
A denoiser trained on discrete one-hot tokens accepts continuous
probability vectors at inference time without retraining: multiplying
row distributions $\mathbf{s} \in (\Delta^{V-1})^L \subset
\mathbb{R}^{L \times V}$ by the token embedding matrix $E \in
\mathbb{R}^{V \times d}$ yields mixture embeddings $\mathbf{s} E \in
\mathbb{R}^{L \times d}$ that reduce to exact token lookups at every
one-hot position. We initialize every non-special position $i \notin
\mathcal{C}$ to the one-hot mask distribution $\mathbf{s}_{t_0}^i =
\delta_m \in \Delta^{V-1}$ (holding special tokens fixed in
$\mathcal{C}$). At step $k$ along $t_k = k/K$, the decoder receives
$\mathbf{s}_{t_k} E$, and each uncrystallized position steps forward
along the velocity field (Equation~\eqref{eq:dfm-velocity}):
\begin{equation}
  \mathbf{s}_{t_{k+1}}^i \;=\;
  \bigl(1 - \tfrac{h}{1-t_k}\bigr)\, \mathbf{s}_{t_k}^i
  \;+\; \tfrac{h}{1-t_k}\,
  p_{1 \mid t_k}^\theta(\cdot \mid \mathbf{s}_{t_k} E, \mathbf{h}_\mathrm{spec})^i,
  \quad i \notin \mathcal{C}.
  \label{eq:soft}
\end{equation}
Position $i$ crystallizes as soon as its maximum-probability token
leaves the mask, $v^\star = \arg\max_v \mathbf{s}_{t_{k+1}}^i(v) \neq
m$: we set $\mathbf{s}_{t_{k+1}}^i \gets \delta_{v^\star}$, add $i$ to
the frozen set $\mathcal{C}$, and leave it unchanged on all remaining
steps. Subsequent passes therefore attend to crisp one-hot embeddings
at crystallized positions and soft mixture embeddings at positions
still being refined. Once $\mathcal{C}$ covers all $L$ positions of a
spectrum, decoding exits early for that spectrum
(Algorithm~\ref{alg:cf}).

\begin{algorithm}[!t]
\caption{Crystallizing Flows inference.}
\label{alg:cf}
\small
\begin{algorithmic}[1]
\Require denoiser $p_{1\mid t}^\theta$, token embedding matrix $E \in \mathbb{R}^{V \times d}$,
         spectrum embedding $\mathbf{h}_\mathrm{spec}$, step budget $K$
\State Initialize $\mathbf{s}^i \gets \delta_m$ for non-special $i \in \{1, \ldots, L\}$
       and one-hot specials on $\mathcal{C} \gets \{\text{special positions}\}$
\State $t \gets 0$;\quad $h \gets 1/K$
\For{$k = 0, \ldots, K - 1$}
  \State $\mathbf{p} \gets
         p_{1 \mid t}^\theta(\cdot \mid \mathbf{s} E, \mathbf{h}_\mathrm{spec})$
         \Comment{1 parallel forward pass; $\mathbf{p}^i \in \Delta^{V-1}$}
  \For{each uncrystallized position $i \notin \mathcal{C}$}
    \State $\mathbf{s}^i \gets
           \bigl(1 - \tfrac{h}{1 - t}\bigr)\,\mathbf{s}^i
           + \tfrac{h}{1 - t}\, \mathbf{p}^i$
           \Comment{simplex Euler step ($i \in \mathcal{C}$ stay frozen)}
    \If{$v^\star \gets \arg\max_v \mathbf{s}^i(v) \neq m$}
      \State $\mathbf{s}^i \gets \delta_{v^\star}$;\quad
             $\mathcal{C} \gets \mathcal{C} \cup \{i\}$
             \Comment{crystallize to one-hot \& freeze}
    \EndIf
  \EndFor
  \If{$\mathcal{C} = \{1, \ldots, L\}$}
    \State \textbf{break}
           \Comment{per-spectrum early exit}
  \EndIf
  \State $t \gets t + h$
\EndFor
\State \Return $X^i = \arg\max_v\, \mathbf{s}^i(v)$ for $i = 1, \ldots, L$
\end{algorithmic}
\end{algorithm}

Appendix~\ref{app:cf-why} details why distribution propagation,
determinism, clean conditioning at decided positions, and midpoint
early exit each contribute to the accuracy gain.

\section{Experiments}\label{sec:methodology}

\paragraph{Dataset, benchmark split, and per-spectrum evaluation.}
All models train on CandyCrunch~v2~\citep{candycrunch2024,zenodo2024cc2} (503,446 annotated LC-MS/MS spectra), filtered to the 497,751 rows (98.9\%) representable in GlycoBART's antenna notation. Even after CandyCrunch~v2 merges consecutive scans within an elution peak, a single LC-MS acquisition records repeated spectra of the same glycan across co-eluting scans, charge states, adduct ions, and separated anomer peaks. Spectrum-level random splitting scatters roughly 85\% of these near-duplicate scans into training and 15\% into testing, turning structure prediction into chromatographic-peak retrieval and inflating reported accuracy---the regime used in~\citet{glycobart2025}. We instead construct a leak-free \textbf{CandyCrunch~v2 benchmark} split by partitioning entire LC-MS files by source filename into 420,977 training and 76,774 test spectra (holding out 10\% of training for validation), and retrain GlycoBART from scratch on this split.

At test time, whereas offline sample-level pipelines aggregate 5 to 50 scans per chromatographic peak, we evaluate single-spectrum prediction directly in both batched and serial streaming regimes, predicting one IUPAC structure per spectrum and scoring exact graph isomorphism via \texttt{glycowork}~\citep{glycowork2021} without post-processing (no multi-scan consensus voting, charge-state merging, or reference-database snap-back). In~\citet{glycobart2025}, combining random splitting with multi-scan voting and database snap-back lifts GlycoBART top-1 accuracy to 89.1\%, gaining 5--15 percentage points from cross-scan aggregation unavailable live (Appendix~\ref{app:glycobart}).

\paragraph{Training and reproducibility.}
We train all headline models on a single 141\,GB H200 GPU in BF16 precision. GlycoGen-DFM, GlycoGen-MDLM, and GlycoGen-AR share the exact same optimization recipe: AdamW~\citep{loshchilov2019adamw} at batch size 1024, base learning rate $3 \times 10^{-4}$ scaled by $\sqrt{8}$ to $8.49 \times 10^{-4}$ over 30 epochs with 200-step linear warmup and cosine decay to 3\% of peak, weight decay 0.01, and gradient clipping at 1.0. GlycoBART follows~\citet{glycobart2025} at batch size 160 ($\sqrt{2}$-scaled learning rate; Appendix~\ref{app:glycobart}). We selected the GlycoGen architecture and schedule via a two-stage \textbf{Distillate} automated research exploration~\citep{lacombe2026distillate} (single-GPU sweeps followed by 21 runs on A100 and L40S GPUs scaling top candidates to 50M parameters on a CandyCrunch subset); GlycoGen-MDLM and GlycoGen-AR reuse this recipe unchanged.

\begin{table}[!t]
\centering
\small
\renewcommand{\arraystretch}{1.06}
\begin{tabular*}{\linewidth}{@{\extracolsep{\fill}}lrrrrr}
\toprule
\textbf{Model (operating point)} & \textbf{NFE} & \textbf{ms/spec} & \textbf{spec/s} & \textbf{Accuracy (3 seeds)} & \textbf{Speedup} \\
\midrule
\multicolumn{6}{l}{\emph{GlycoGen-DFM $+$ Crystallizing Flows (ours), $50$M}} \\
\quad $K{=}1$ (one-pass)              & $1$       & $0.046$ & $21{,}883$ & $\mathbf{0.9019 \pm 0.0004}$  & $\mathbf{61\times}$ \\
\quad $K{=}8$ (recommended)           & $5$       & $0.201$ & $4{,}975$  & $\mathbf{0.9251 \pm 0.0006}$  & $\mathbf{13.9\times}$ \\
\quad $K{=}32$                        & $17$      & $0.675$ & $1{,}481$  & $\mathbf{0.9267 \pm 0.0005}$  & $\mathbf{4.1\times}$ \\
\quad $K{=}64$                        & $34$      & $1.306$ & $766$      & $\mathbf{0.9269 \pm 0.0006}$  & $\mathbf{2.1\times}$ \\
\midrule
\multicolumn{6}{l}{\emph{GlycoGen-MDLM $+$ Crystallizing Flows (ours, cross-objective control), $50$M}} \\
\quad $K{=}1$ (one-pass)              & $1$       & $0.046$ & $21{,}739$ & $0.8973 \pm 0.0017$           & $61\times$ \\
\quad $K{=}8$                         & $5$       & $0.199$ & $5{,}025$  & $0.9230 \pm 0.0012$           & $14.1\times$ \\
\quad $K{=}32$                        & $17$      & $0.676$ & $1{,}479$  & $0.9247 \pm 0.0011$           & $4.1\times$ \\
\midrule
\multicolumn{6}{l}{\emph{GlycoGen-AR (matched-architecture autoregressive control), $50$M}} \\
\quad beam-$1$ (top-$1$)              & $33$      & $2.21$  & $452$      & $0.9038 \pm 0.0005$           & $1.3\times$ \\
\quad beam-$32$ (top-$1$, seed $0$)   & $1{,}045$ & $72.66$ & $14$       & $0.9039$                      & $0.04\times$ \\
\midrule
\multicolumn{6}{l}{\emph{GlycoBART reproduction~\citep{glycobart2025}, $207$M (per-spectrum)}} \\
\quad beam-$1$ (top-$1$)              & $23$      & $2.80$  & $357$      & $0.8092 \pm 0.0031$           & $1.0\times$ \\
\quad beam-$8$ (top-$5$, seed $0$)    & $180$     & $27.12$ & $37$       & $0.8578$                      & $0.10\times$ \\
\quad beam-$32$ (top-$5$, seed $0$)   & $720$     & $91.21$ & $11$       & $0.8635$                      & $0.03\times$ \\
\bottomrule
\end{tabular*}
\vspace{0.35em}
\caption{\textbf{Headline results.} Top-1 per-spectrum structural
accuracy on the 76,774 test spectra of the CandyCrunch~v2 benchmark
(mean $\pm$ SD over three seeds on one H200 GPU in BF16; full sweep in
Table~\ref{tab:multiseed-sweep}; high-beam sweeps on seed 0 as top-1 is
flat on AR and drops above beam 1 on GlycoBART). Batched speedup
compares against greedy GlycoBART ($2.80$\,ms/spec $= 1.0\times$);
relative to beam-32 ($91.21$\,ms/spec), 8-step GlycoGen is $454\times$
faster. Unbatched serial streaming latencies appear in
Table~\ref{tab:streaming-latency}.}
\label{tab:results}
\end{table}

\paragraph{Models and baselines.}
We compare four models on identical splits ($K \in \{1,2,4,8,16,32,64\}$ for flow/diffusion; beam widths $1\text{--}32$ for autoregressive baselines):
\begin{itemize}\setlength{\itemsep}{1pt}\setlength{\parskip}{0pt}\setlength{\topsep}{2pt}
\item \textbf{GlycoGen-DFM + Crystallizing Flows:} our 50M-parameter encoder-decoder DFM model (Section~\ref{sec:arch}), trained with uniform continuous time sampling and a linear schedule.
\item \textbf{GlycoGen-MDLM + Crystallizing Flows:} a matched 50M MDLM denoiser~\citep{sahoo2024mdlm} trained on a 128-step grid with NELBO cross-entropy (Appendix~\ref{app:mdlm}).
\item \textbf{GlycoGen-AR:} a matched 46.4M causal autoregressive control trained with teacher-forced cross-entropy at beams 1--32 ($\sim$33 passes/beam; Appendix~\ref{app:ar-baseline}).
\item \textbf{GlycoBART reproduction:} the 207M autoregressive model of~\citet{glycobart2025}, retrained on our CandyCrunch~v2 benchmark split. Unlike GlycoGen's atomic IUPAC vocabulary, GlycoBART predicts compressed antenna motifs whose reconstruction back to full IUPAC trees loses 10.3 percentage points of top-1 accuracy from branch-attachment ambiguities (Appendix~\ref{app:glycobart}) and excludes curly-brace strings (\texttt{\{...\}}, 1.1\% of CandyCrunch~v2).
\end{itemize}

\section{Results}\label{sec:results}

\begin{figure}[!t]
\centering
\includegraphics[width=\textwidth]{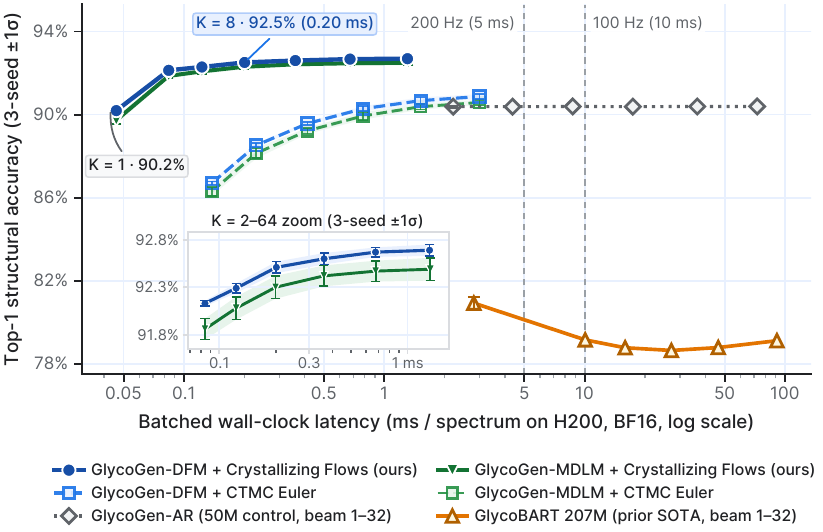}
\caption{\textbf{Crystallizing Flows Pareto-dominate at low step
counts on both accuracy and wall-clock cost.} Top-1 per-spectrum
structural accuracy on the 76,774 test spectra of the CandyCrunch~v2
benchmark (mean $\pm 1\sigma$ over three seeds; inset zooms
into $K{=}2\text{--}64$) versus batched latency on one H200 GPU. DFM
with Crystallizing Flows (\textcolor[HTML]{174EA6}{\textbf{deep blue}},
ours) and MDLM with Crystallizing Flows
(\textcolor[HTML]{137333}{\textbf{green}}, ours) Pareto-dominate CTMC
Euler, 50M GlycoGen-AR, and 207M GlycoBART~\citep{glycobart2025}.
Dashed lines mark 200\,Hz ($5$\,ms) and 100\,Hz ($10$\,ms) scan
intervals (Table~\ref{tab:streaming-latency}).}
\label{fig:pareto-ms}
\end{figure}

\paragraph{Headline accuracy and ablation of design choices.}
Across all 76,774 test spectra in the CandyCrunch~v2 benchmark, single-pass GlycoGen-DFM reaches \textbf{90.19\%} top-1 structural accuracy, already surpassing every autoregressive baseline. Refining predictions with Crystallizing Flows lifts accuracy to \textbf{92.51\%} at 5 effective passes ($K{=}8$), \textbf{92.60\%} at 9 passes ($K{=}16$), \textbf{92.67\%} at 17 passes ($K{=}32$), and \textbf{92.69\%} at 34 passes ($K{=}64$) across three seeds (Table~\ref{tab:results}, Figure~\ref{fig:pareto-ms}). Comparing controlled baselines isolates where this 11.7 percentage point improvement over greedy GlycoBART (80.92\%) comes from:
\begin{enumerate}\setlength{\itemsep}{1pt}\setlength{\parskip}{0pt}\setlength{\topsep}{2pt}
\item \textbf{Full-spectrum patch encoding and direct IUPAC tokenization (+9.5 pp).} Instead of truncating spectra to top-$N$ discrete peak tokens and predicting lossy antenna strings, encoding the full spectrum as 64 patches of binned intensities and sub-bin mass offsets while generating atomic IUPAC tokens directly lifts GlycoGen-AR from 80.92\% to 90.38\%, preserving weak diagnostic fragments and eliminating the $\sim$9\% antenna loss.
\item \textbf{Parallel decoding with flow matching and masked diffusion (+2.1 pp).} Decoding all sequence positions simultaneously rather than one token at a time from left to right lets every position attend bidirectionally to both sides of a branched glycan tree as structure resolves, lifting accuracy from 90.38\% on GlycoGen-AR to 92.51\% on GlycoGen-DFM at $K{=}8$.
\item \textbf{Crystallizing Flows (+1.8 to +5.6 pp over Euler).} Propagating full token probability distributions on the simplex instead of drawing random categorical commits adds 5.43 percentage points over standard CTMC Euler at $K{=}2$ (5.56 pp on MDLM) and 1.82 percentage points at $K{=}64$ (1.91 pp on MDLM).
\end{enumerate}

\paragraph{Inference speed.}
Decoding all positions in parallel rather than stepping through $\sim$33 autoregressive passes per beam makes GlycoGen one to two orders of magnitude faster than GlycoBART in both batch processing and serial streaming. We benchmarked all models on an identical H200 VM on Hugging Face Jobs (Tables~\ref{tab:results} and~\ref{tab:streaming-latency}); as absolute timings vary across GPU generations and serving stacks, these numbers provide one controlled reference point. On this setup, batched GlycoGen runs $61\times$ faster than greedy GlycoBART at $K{=}1$ and $13.9\times$ faster at $K{=}8$ ($454\times$ faster than beam-32), turning multi-day repository runs into minutes. In unbatched serial streaming, end-to-end latency including CPU peak binning drops from $183$\,ms/spec for greedy GlycoBART to $8.6$\,ms at $K{=}1$ (116\,Hz, matching 100\,Hz scan cadence) and $33.8$\,ms at $K{=}8$.

\paragraph{Comparison with alternative samplers.}
Crystallizing Flows outperform standard CTMC Euler at every step count from $K{=}2$ to $K{=}64$ (Figure~\ref{fig:pareto-ms}). At $K{=}2$, CTMC Euler drops to 86.70\%---3.5 percentage points below a single pass---as random categorical commits lock in early errors. Across seven other discrete and simplex samplers on the same checkpoint (temperature-scaled, predictor-corrector, pure soft-propagation, and confidence-ordered unmasking), Crystallizing Flows outperform every alternative at low step budgets, with confidence unmasking matching accuracy only above 33 passes (Appendix~\ref{app:samplers}).

\paragraph{Cross-instrument robustness and generalization to unseen IUPAC spellings.}
Mass spectrometers vary widely in collision energetics and mass resolution, yet single-pass GlycoGen maintains 88.3\% to 93.0\% top-1 accuracy across all four analyzer families in CandyCrunch~v2 (linear ion traps, Orbitraps, Bruker amaZon, and other analyzers; Table~\ref{tab:cross-instrument}).

We also tested whether GlycoGen learns chemical graph structure from string supervision or merely memorizes literal training sequences. Because a glycan is a branched tree rather than a linear chain, a single molecule can be written as several distinct valid IUPAC strings depending on which branch is serialized first (Figure~\ref{fig:qual-snfg}, Appendix~\ref{app:inst-gen}). Grouping the 76,774 test spectra by chemical graph isomorphism via \texttt{glycowork}~\citep{glycowork2021} yields 76,203 spectra matching a training string verbatim, 66 spectra whose true glycan graph appears in training \emph{only} under a different branch ordering, and 505 spectra with completely unseen glycan topologies (Table~\ref{tab:struct-gen}). On those 66 unseen-spelling spectra, GlycoBART drops to 13.6\%, failing to recognize known molecules written in a new branch order. Direct IUPAC tokenization lifts GlycoGen-AR to 54.5\%, and parallel bidirectional decoding lifts GlycoGen further to \textbf{60.6\%} (DFM) and \textbf{62.1\%} (MDLM)---clear evidence that \textbf{GlycoGen generalizes across unseen string representations of known chemical graphs}.

\paragraph{Transfer to masked diffusion, early exit, and autoregressive beam saturation.}
Applied without retraining as a drop-in sampler for our 128-step MDLM denoiser~\citep{sahoo2024mdlm}, Crystallizing Flows improve over MDLM's CTMC Euler sampler by 5.56 percentage points at $K{=}2$ and 1.91 percentage points at $K{=}64$, tracking GlycoGen-DFM within 0.21 percentage points from $K{=}4$ upward (Appendix~\ref{app:mdlm}). Meanwhile, per-spectrum early exit cuts neural forward passes nearly in half ($17.4$ passes at $K{=}32$ and $33.9$ at $K{=}64$). Widening beam search on GlycoGen-AR from beam 1 to beam 32 leaves top-1 accuracy flat at $90.38\% \pm 0.05\%$ despite a $33\times$ latency penalty, showing that wider left-to-right search cannot substitute for iterative parallel refinement.

\paragraph{What explains the remaining errors?}
On the 76,203 test spectra whose target structure appears in training, GlycoGen reaches 93.31\% exact graph accuracy (Table~\ref{tab:struct-gen}), and 99.1\% of its 5,627 mismatched predictions at $K{=}32$ are syntactically valid IUPAC strings (over 92\% parse into complete chemical graphs in \texttt{glycowork}). Roughly four in five valid mismatches reflect physical limits of single-spectrum CID rather than hallucinations (Table~\ref{tab:qual-examples}, Figure~\ref{fig:qual-snfg}): 40--42\% match monosaccharide composition and differ only in regioisomer linkage ($\beta 1\text{-}3$ vs.\ $\beta 1\text{-}4$) or linkage wildcard specificity (\texttt{a2-?}), which share glycosidic cleavage masses absent cross-ring ions, while $\sim$38\% differ by one labile fucose or sialic acid lost in-source before MS/MS, approaching the Bayes error floor of low-energy CID.

\section{Limitations and Future Work}\label{sec:discussion}

\subsection{Limitations}\label{sec:limit}

\paragraph{Generalization to novel glycan topologies.}
On the 505 test spectra whose chemical graph never appears in training, all four models score near zero ($0.2\%$--$3.6\%$; Table~\ref{tab:struct-gen}). First, CandyCrunch~v2 concentrates roughly 500,000 spectra onto 3,590 unique glycan graphs, limiting structural breadth. Second, fixed-$m/z$ bin encoding favors whole-spectrum pattern recognition over translation-invariant $\Delta m/z$ reasoning when unfamiliar branches or reducing-end tags shift fragments across bins.

\paragraph{Zero-shot transfer across mass analyzer platforms.}
While GlycoGen reaches 88.3\%--93.0\% accuracy across all four analyzer families when present in training, holding out an entire family lowers zero-shot accuracy to 41.4\% on linear ion traps and 14.1\% on Orbitraps (Table~\ref{tab:cross-instrument}); distinct collision energetics mean new platforms (e.g., timsTOF, Orbitrap Astral) require fine-tuning.

\paragraph{Notation scope and crystallization calibration.}
For direct comparison with GlycoBART, we exclude the 1.1\% of CandyCrunch~v2 spectra with curly-brace substituents (\texttt{\{...\}}), though atomic IUPAC tokens represent them directly. Under $\kappa_t = t$, positions crystallize at mean probability $0.56 \pm 0.06$ at $K{=}8$ (Appendix~\ref{app:inst-gen}), calibrated to linear interpolation; delayed schedules can use a threshold.

\subsection{Future Work}\label{sec:future}

\paragraph{Generalizing to novel glycan structures.}
Moving from recognition of known topologies to true \emph{de novo} discovery requires building fragmentation physics into both encoder and training data. Replacing fixed-$m/z$ patch embeddings with pairwise peak-difference ($\Delta m/z$) attention makes neutral-loss reasoning invariant to precursor mass shifts and reducing-end tags, while pretraining on simulated fragmentation trees across GlyTouCan~\citep{tiemeyer2017glytoucan} can expose the model to tens of thousands of topologies absent from current benchmarks. Pairing compositional pretraining with calibrated simplex probabilities would also let the decoder emit explicit linkage wildcards (\texttt{a2-?}) when CID leaves regioisomers underdetermined.

\paragraph{Illuminating the dark glycome across large-scale experimental datasets.}
Public glycomics and glycoproteomics archives hold tens of millions of raw LC-MS/MS spectra across tissues, species, and disease cohorts that remain unannotated---a ``dark glycome'' unexplored for lack of a method both fast enough for repository throughput and general enough for open-vocabulary prediction. Sub-millisecond batched decoding with direct IUPAC generation makes exhaustive reanalysis practical on one GPU, turning unannotated repositories into systematic surveys of glycosylation diversity across species and disease.

\section{Conclusion}\label{sec:conclusion}

We introduced \textbf{GlycoGen} and \textbf{Crystallizing Flows} for open-vocabulary glycan structure prediction from tandem mass spectra. Across the CandyCrunch~v2 benchmark, GlycoGen reaches 92.67\% top-1 accuracy---93.31\% on seen structures, approaching the single-spectrum CID Bayes error ceiling---while generalizing to unseen IUPAC branch orderings, transferring without retraining to masked diffusion, and accelerating inference by one to two orders of magnitude. Methodologically, these results show that \emph{de novo} elucidation of branched molecules is poorly served by left-to-right token search: a fragment spectrum probes the entire glycan tree at once, and decoding all positions simultaneously on the probability simplex is both faster and more faithful to the underlying chemistry. Biologically, whereas genomics and proteomics scaled once sequences became routine to read, glycomics has remained bottlenecked by slow, library-bound spectrum interpretation. Making the glycome readable directly from mass spectra in real time and at scale removes a central barrier to mapping how glycosylation governs cellular signaling, immunity, and human health.

\paragraph{Code and data availability.}
Open-source code, trained three-seed checkpoints, and our CandyCrunch~v2 benchmark splits are available at \url{https://github.com/rlacombe/GlycoGen} and \url{https://huggingface.co/datasets/rlacombe/candycrunch-v2-benchmark}.

\clearpage
\begin{ack}
The author thanks Emilia McLaughlin and Marina Lochhead (Kyron Bio) for invaluable discussions on experimental glycobiology workflows that motivated this work, the CandyCrunch and GlycoBART authors for open-sourcing the datasets and pipelines on which it builds, and the maintainers of \texttt{glycowork} for the graph-isomorphism evaluation library used throughout. This research was independently self-funded by the author, including LLM token budgets for Distillate auto-research explorations, as well as training and evaluation runs on Hugging Face Jobs GPUs.
\end{ack}

\bibliographystyle{plainnat}
\bibliography{references}

\appendix

\section{Mask-source CTMC Euler update}\label{app:alg1}

For reference, Algorithm~\ref{alg:euler} states the standard CTMC Euler
update for mask-source DFM (the mask-source specialization of
Algorithm~1 of \citealp{gat2024dfm}). At the final step $k = K - 1$,
$t = 1 - 1/K$ so $h / (1 - t) = 1$, committing every remaining masked
position within $K$ forward passes.

\begin{algorithm}[H]
\caption{CTMC Euler update for mask-source DFM (\citealp{gat2024dfm}, Algorithm 1).}
\label{alg:euler}
\small
\begin{algorithmic}[1]
\Require denoiser $p_{1 \mid t}^\theta$, spectrum embedding $\mathbf{h}_\mathrm{spec}$,
         step budget $K$
\State Initialize $X^i \gets m$ for non-special $i \in \{1, \ldots, L\}$
       and fixed specials on $\mathcal{C} \gets \{\text{special positions}\}$
\State $t \gets 0$;\quad $h \gets 1/K$
\For{$k = 0, 1, \ldots, K - 1$}
  \State $\mathbf{p} \gets p_{1 \mid t}^\theta(\cdot \mid X, \mathbf{h}_\mathrm{spec})$
         \Comment{1 parallel forward pass; $\mathbf{p}^i \in \Delta^{V-1}$}
  \For{each masked position $i \notin \mathcal{C}$ with $X^i = m$}
    \State $r \sim \mathrm{Uniform}(0, 1)$
    \If{$r < h / (1 - t)$}
      \State $X^i \sim \mathbf{p}^i$
      \Comment{commit a sampled token and freeze}
    \EndIf
  \EndFor
  \State $t \gets t + h$
\EndFor
\State \Return $X$
\end{algorithmic}
\end{algorithm}

\section{Spectrum dual-channel binning and patch encoding}\label{app:patching}

\begin{figure}[!t]
\centering
\includegraphics[width=0.94\textwidth]{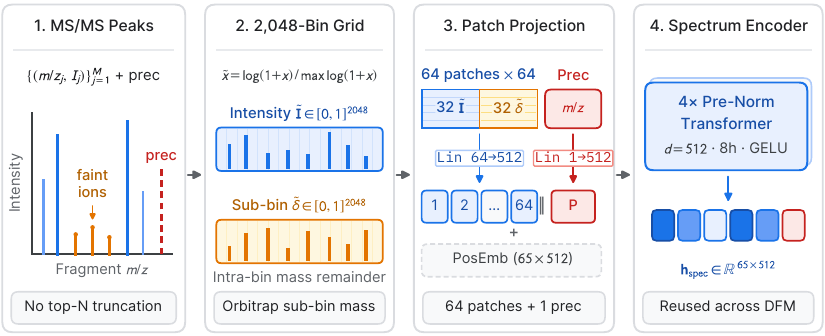}
\caption{\textbf{Dual-channel spectrum binning and patch encoding.}
Variable-length fragment peak lists $\{(m/z_j, I_j)\}_{j=1}^M$ map
without top-$N$ peak truncation onto a 2,048-bin mass-to-charge grid
carrying two normalized channels per bin: log-scaled peak intensity
$\tilde{\mathbf{I}} \in [0, 1]^{2048}$ and continuous sub-bin mass
remainder $\tilde{\boldsymbol{\delta}} \in [0, 1]^{2048}$. Reshaping
into 64 contiguous patches of 32 bins ($64$ features per patch),
projecting to $d = 512$, and appending one standardized precursor-$m/z$
token yields a fixed 65-token sequence encoded once per spectrum into
$\mathbf{h}_\mathrm{spec} \in \mathbb{R}^{65 \times 512}$.}
\label{fig:spectrum-patching}
\end{figure}

Raw tandem mass spectra vary widely in peak density, and low-intensity
fragment ions frequently carry diagnostic branching or linkage
information that top-$N$ peak filtering discards. Following the binned
CandyCrunch~v2 representation~\citep{candycrunch2024}, we discretize
each spectrum across $B_\mathrm{mz} = 2{,}048$ uniform bins spanning the
glycan fragment mass window alongside a continuous sub-bin mass-offset
vector recording the exact intra-bin $m/z$ remainder of peaks falling
in each bin. We normalize both vectors
$\mathbf{x} \in \mathbb{R}_{\ge 0}^{2048}$ per spectrum via
$\tilde{\mathbf{x}} = \log(1 + \mathbf{x}) / (\max_b \log(1 + x_b) +
10^{-6})$, compressing dynamic range while preserving faint peaks and
Orbitrap-grade mass offsets.

Rather than feeding 2,048 individual bin tokens into quadratic
self-attention, the spectrum encoder groups every $C = 32$ contiguous
bins into $N_\mathrm{patch} = 64$ non-overlapping patches and
concatenates the intensity and sub-bin offset channels within each
patch to form a $64 \times 64$ patch matrix
(Figure~\ref{fig:spectrum-patching}). A linear layer
$\mathbb{R}^{64} \to \mathbb{R}^{512}$ projects the 64 patches into
model width $d = 512$, while a separate linear map
$\mathbb{R}^1 \to \mathbb{R}^{512}$ projects the scalar precursor $m/z$
(z-score standardized using training-set mean and standard deviation)
as token 65. Adding learned positional embeddings in
$\mathbb{R}^{65 \times 512}$ and applying 4 pre-norm GELU Transformer
encoder layers ($8$ heads, feed-forward width $2{,}048$) computes
$\mathbf{h}_\mathrm{spec} \in \mathbb{R}^{65 \times 512}$ in a single
encoder pass cached across all $K$ decoder steps.

\section{Crystallizing Flows: mechanistic intuitions}\label{app:cf-why}

\textbf{Distribution propagation preserves information.} The
per-position softmax over the 74-token vocabulary (70 atomic IUPAC
tokens plus 4 special tokens) carries up to $\log_2 74 \approx 6.2$
bits of information, compared to zero bits once a token commits to a
single value. CTMC Euler discards this distribution at every commit step and
replaces a calibrated belief with a single Bernoulli draw; once committed,
positions remain frozen and cannot recover lost probability mass.
Crystallizing Flows instead preserve the full simplex state until the
model's own argmax leaves \texttt{<mask>}, collapsing only positions
on which the denoiser has formed a clear consensus.

\textbf{Determinism eliminates sampling variance.} CTMC Euler injects
per-step Bernoulli noise at rate $h/(1-t)$. Sampling variance compounds
across positions and steps; at small step budgets $K$, the trajectory
drifts far from the posterior mean and accuracy regresses below a single
forward pass, dropping 3.5 percentage points to 86.70\% across three seeds
at $K{=}2$ (86.76\% on seed 0; Tables~\ref{tab:multiseed-sweep}
and~\ref{tab:samplers}). Crystallizing Flows take the closed-form
posterior-mean update at every step without random draws, improving
accuracy monotonically with $K$.

\textbf{Clean conditioning at decided positions sharpens remaining
positions.} Crystallized positions feed the decoder discrete one-hot
tokens, while uncrystallized positions feed soft mixture embeddings.
Trained on discrete one-hot inputs, the decoder treats crystallized
positions as observed context and concentrates attention on resolving the
remaining uncertain positions, tightening the conditional sub-problem on
each commit.

\textbf{Why early exit averages roughly half the step budget.} Under
the linear schedule $\kappa_t = t$, a trajectory starting at
$\mathbf{s}_0^i = \delta_m$ whose denoiser places zero mass on
\texttt{<mask>} retains exact mask mass $\mathbf{s}_{t_k}^i(m) = 1 - t_k$
at step $t_k = k/K$. Because $\mathbf{s}_t^i(m) \ge 0.5$ for all
$t \le 0.5$, no position can leave \texttt{<mask>} before the midpoint.
On the first step past $t = 0.5$ ($t_{k+1} = 0.5 + 1/K$, forward pass
$\lfloor K/2 \rfloor + 1$), mask mass drops below $0.5$ and every
confident position whose leading token exceeds $1 - t_{k+1}$
crystallizes immediately ($91.6\%$ of commits at $K{=}8$ occur on pass 4
and the remaining $8.4\%$ on pass 5; Section~\ref{app:inst-gen}). Across
all 76,774 test spectra, decoding therefore averages roughly $K/2 + 1$
effective passes, delivering measured wall-clock speedups from
$1.28\times$ at $K{=}4$ to $1.84\times$ at $K{=}64$.

\textbf{Why no unmasking schedule or temperature is needed.}
MaskGIT-style top-$N$ schedulers require choosing how many positions to
commit per step, and temperature samplers require tuning a softmax
temperature. Crystallizing Flows require neither: commits trigger
directly when a position's simplex argmax overtakes \texttt{<mask>},
calibrating unmasking timing automatically from the flow schedule.

\section{Multi-seed replication, generalization, and qualitative error analysis}\label{app:inst-gen}

\begin{figure}[!t]
\centering
\includegraphics[width=0.90\textwidth]{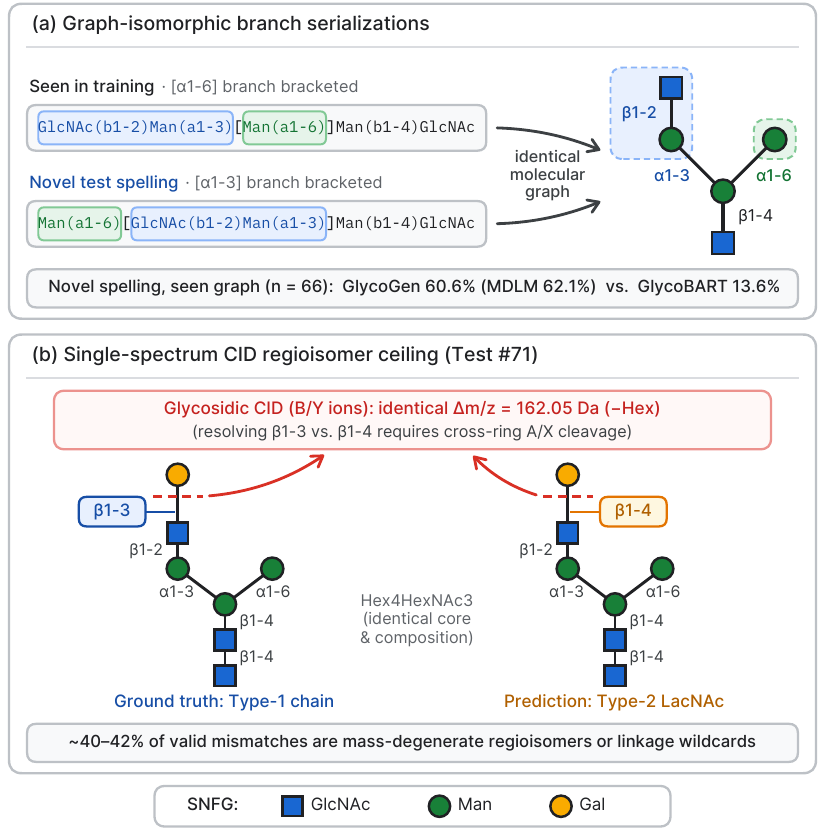}
\caption{\textbf{Graph-isomorphic generalization and the single-spectrum CID regioisomer ceiling.}
\textbf{(a)} Distinct IUPAC bracket traversal orders serialize the exact same branched SNFG molecular graph; GlycoGen generalizes across unseen branch spellings ($60.6\%$ DFM, $62.1\%$ MDLM vs.\ $13.6\%$ GlycoBART on $n{=}66$).
\textbf{(b)} Terminal $\beta 1\text{-}3$ vs.\ $\beta 1\text{-}4$ galactose regioisomers (\texttt{\#71}) yield identical glycosidic CID neutral losses ($\Delta m/z = 162.05$\,Da), illustrating the physical single-spectrum CID ambiguity behind $\sim$40--42\% of valid mismatches.}
\label{fig:qual-snfg}
\end{figure}

\paragraph{Full three-seed replication across step budgets.}
Table~\ref{tab:multiseed-sweep} reports the three-seed replication
(initialization seeds 0, 1, and 2) across step budgets $K \in \{1, 2,
4, 8, 16, 32, 64\}$ for GlycoGen-DFM and MDLM under both standard
CTMC Euler and Crystallizing Flows, matching the curves and $\pm 1\sigma$
bands in Figure~\ref{fig:pareto-ms}.

\begin{table}[ht]
\centering
\small
\setlength{\tabcolsep}{2.5pt}
\begin{tabular*}{\linewidth}{@{\extracolsep{\fill}}rccccc}
\toprule
\textbf{$K$} & \textbf{Eff.\ NFE (CF)} & \textbf{DFM + CF} & \textbf{DFM + CTMC Euler} & \textbf{MDLM + CF} & \textbf{MDLM + CTMC Euler} \\
\midrule
$1$  & $1.00$  & $0.9019 \pm 0.0004$ & $0.9019 \pm 0.0004$ & $0.8973 \pm 0.0017$ & $0.8973 \pm 0.0017$ \\
$2$  & $2.00$  & $0.9213 \pm 0.0003$ & $0.8670 \pm 0.0008$ & $0.9186 \pm 0.0011$ & $0.8630 \pm 0.0021$ \\
$4$  & $2.99$  & $0.9229 \pm 0.0005$ & $0.8852 \pm 0.0010$ & $0.9208 \pm 0.0012$ & $0.8814 \pm 0.0025$ \\
$8$  & $5.05$  & $0.9251 \pm 0.0006$ & $0.8957 \pm 0.0008$ & $0.9230 \pm 0.0012$ & $0.8922 \pm 0.0015$ \\
$16$ & $9.16$  & $0.9260 \pm 0.0006$ & $0.9028 \pm 0.0012$ & $0.9242 \pm 0.0011$ & $0.8993 \pm 0.0018$ \\
$32$ & $17.40$ & $0.9267 \pm 0.0005$ & $0.9067 \pm 0.0016$ & $0.9247 \pm 0.0011$ & $0.9038 \pm 0.0014$ \\
$64$ & $33.89$ & $0.9269 \pm 0.0006$ & $0.9087 \pm 0.0011$ & $0.9249 \pm 0.0012$ & $0.9058 \pm 0.0013$ \\
\bottomrule
\end{tabular*}
\vspace{0.5em}
\caption{Three-seed top-1 structural accuracy (mean $\pm$ standard deviation over initialization seeds 0, 1, and 2) across the 76,774 spectra of the CandyCrunch~v2 benchmark test set. Eff.\ NFE (CF) reports mean forward passes under Crystallizing Flows early exit; CTMC Euler evaluates all $K$ passes.}
\label{tab:multiseed-sweep}
\end{table}

\paragraph{Structure-level generalization breakdown.}
Table~\ref{tab:struct-gen} splits the 76,774 test spectra into three
mutually exclusive regimes by comparing each target against the
training set via exact IUPAC string matching and \texttt{glycowork}
graph isomorphism:
76,203 spectra whose exact IUPAC string appears in training, 66
spectra whose glycan graph appears in training only under a different
IUPAC branch ordering, and 505 spectra whose glycan graph is absent
from training under any string encoding.

\begin{table}[ht]
\centering
\small
\setlength{\tabcolsep}{1.5pt}
\begin{tabular*}{\linewidth}{@{\extracolsep{\fill}}lccc}
\toprule
\textbf{Model} & \textbf{Seen str.\ ($76{,}203$)} & \textbf{Novel str., seen graph ($66$)} & \textbf{Novel graph ($505$)} \\
\midrule
GlycoGen-DFM ($K{=}1$)        & $0.9081 \pm 0.0004$          & $0.6061$ $[0.485, 0.715]$          & $0.0079$ $[0.003, 0.020]$ \\
GlycoGen-DFM + CF ($K{=}32$)  & $\mathbf{0.9331 \pm 0.0005}$ & $0.5758$ $[0.456, 0.688]$          & $0.0178$ $[0.009, 0.033]$ \\
GlycoGen-MDLM + CF ($K{=}32$) & $0.9311 \pm 0.0011$          & $\mathbf{0.6212}$ $[0.500, 0.729]$ & $0.0020$ $[0.000, 0.011]$ \\
GlycoGen-AR (beam-$1$)        & $0.9102 \pm 0.0005$          & $0.5455$ $[0.426, 0.660]$          & $0.0059$ $[0.002, 0.017]$ \\
GlycoBART (beam-$1$, 3 seeds) & $0.8152 \pm 0.0031$          & $0.1364$ $[0.073, 0.239]$          & $\mathbf{0.0356}$ $[0.023, 0.056]$ \\
\bottomrule
\end{tabular*}
\vspace{0.5em}
\caption{Top-1 structural accuracy disaggregated by target novelty on the 76,774 spectra of the CandyCrunch~v2 benchmark test set. Bracketed intervals on smaller subsets denote 95\% Wilson confidence intervals.}
\label{tab:struct-gen}
\end{table}

\paragraph{Cross-instrument robustness and held-out analyzer retraining.}
Table~\ref{tab:cross-instrument} reports per-instrument accuracy on
the 76,774 test spectra using the true analyzer metadata counts in
$X_\text{test}$, alongside leave-one-instrument-out retraining runs in
which all linear ion trap spectra (220,762 training rows) or all
Orbitrap spectra (113,200 training rows) are excluded from training
and validation and evaluated zero-shot.

\begin{table}[ht]
\centering
\small
\setlength{\tabcolsep}{4pt}
\begin{tabular*}{\linewidth}{@{\extracolsep{\fill}}lrrrr}
\toprule
\textbf{Instrument family (\texttt{trap})} & \textbf{Test $n$} & \textbf{In-dist.\ ($K{=}1$)} & \textbf{In-dist.\ ($K{=}8$ CF)} & \textbf{Held-out 0-shot ($K{=}1$)} \\
\midrule
Linear ion trap (\texttt{linear}) & $39{,}446$ & $0.8829$ & $0.9061$ & $0.414$ $[0.409, 0.419]$ \\
Orbitrap (\texttt{orbitrap})      & $18{,}055$ & $0.9299$ & $0.9508$ & $0.141$ $[0.136, 0.146]$ \\
Bruker amaZon (\texttt{amazon})   & $13{,}046$ & $0.9291$ & $0.9525$ & --- \\
Other (\texttt{other})            & $6{,}227$  & $0.8890$ & $0.9074$ & --- \\
\bottomrule
\end{tabular*}
\vspace{0.5em}
\caption{Per-instrument accuracy on $X_\text{test}$ ($n{=}76{,}774$) and zero-shot leave-one-analyzer-out retraining on linear ion trap and Orbitrap (95\% Wilson CIs). Matched MDLM zero-shot retraining scores $0.404$ on held-out linear ion trap and $0.157$ on held-out Orbitrap.}
\label{tab:cross-instrument}
\end{table}

\paragraph{Pre-commit confidence calibration and qualitative examples.}
Instrumenting all $7{,}084{,}712$ token-crystallization events across
$X_\text{test}$ at $K{=}8$ confirms how Crystallizing Flows differ
from one-shot decoding: whereas single-pass ($K{=}1$) max-softmax
confidences are bimodal (mean $0.739$, median $0.910$, with $33.8\%$
above $0.99$ and $34.1\%$ below $0.50$), multi-step simplex commits
fire as soon as consensus emerges off \texttt{<mask>} (mean pre-commit
simplex probability $0.562 \pm 0.060$, median $0.617$, bounded above at $0.625$
by the linear Euler step; $91.6\%$ of commits occur at step 4 and
$8.4\%$ at step 5). Figure~\ref{fig:qual-snfg} and
Table~\ref{tab:qual-examples} illustrate representative test-set
predictions, graph-isomorphic generalization across branch orderings,
and the physical CID regioisomer ceiling.

\begin{table}[ht]
\centering
\scriptsize
\setlength{\tabcolsep}{3pt}
\renewcommand{\arraystretch}{1.22}
\begin{tabular*}{\linewidth}{@{\extracolsep{\fill}}p{0.23\linewidth}p{0.36\linewidth}p{0.36\linewidth}}
\toprule
\textbf{Regime ($X_\text{test}$ idx)} & \textbf{Ground-truth IUPAC} & \textbf{GlycoGen + CF prediction} \\
\midrule
Exact branched hit (\texttt{\#15964}) &
{\raggedright\texttt{Neu5Ac(a2-?)Gal(b1-4)GlcNAc(b1-2)} \texttt{Man(a1-3)[Man(a1-6)]Man(b1-4)} \texttt{GlcNAc(b1-4)[Fuc(a1-6)]GlcNAc}\par} &
{\raggedright Exact match (monosialylated, core-fucosylated N-glycan).\par} \\
Novel spelling, seen graph (\texttt{\#10631}) &
{\raggedright\texttt{GlcNAc(b1-2)Man(a1-6)[Man(a1-3)]} \texttt{[GlcNAc(b1-4)]Man(b1-4)} \texttt{GlcNAc(b1-4)GlcNAc}\par} &
{\raggedright\texttt{GlcNAc(b1-2)Man(a1-?)[GlcNAc(b1-4)]} \texttt{[Man(a1-?)]Man(b1-4)GlcNAc(b1-4)GlcNAc} (isomorphic bisecting graph).\par} \\
Regioisomer / wildcard (\texttt{\#71}) &
{\raggedright\texttt{Gal(b1-3)GlcNAc(b1-2)Man(a1-3)} \texttt{[Man(a1-6)]Man(b1-4)} \texttt{GlcNAc(b1-4)GlcNAc}\par} &
{\raggedright\texttt{Gal(b1-4)GlcNAc(b1-2)Man(a1-?)} \texttt{[Man(a1-?)]Man(b1-4)GlcNAc(b1-4)GlcNAc} ($\text{Hex}_4\text{HexNAc}_3$ regioisomer).\par} \\
Unseen graph fallback (\texttt{\#72}) &
{\raggedright\texttt{GlcNAc(b1-2)Man(a1-3)[ManOMe(a1-6)]} \texttt{Man(b1-4)GlcNAc(b1-4)GlcNAc}\par} &
{\raggedright\texttt{Fuc(a1-2)Gal(b1-4)GlcNAc(b1-6)} \texttt{[Neu5Ac(a2-3)Gal(b1-3)]GalNAc} (rare \texttt{ManOMe} $\to$ common motif fallback).\par} \\
\bottomrule
\end{tabular*}
\vspace{0.4em}
\caption{Qualitative examples from $X_\text{test}$ illustrating exact branched recovery, graph-isomorphic generalization across branch orderings, single-spectrum CID regioisomer ambiguity, and unseen-topology prior fallback.}
\label{tab:qual-examples}
\end{table}

\section{GlycoBART reproduction and per-spectrum evaluation}\label{app:glycobart}

\paragraph{Setup.}
We reproduce the 207-million-parameter GlycoBART architecture from
\citet{glycobart2025} (12 encoder layers, 12 decoder layers, hidden
width 768, 16 attention heads, and feed-forward dimension 3072) and
retrain it from scratch on a run-stratified partition of
CandyCrunch~\citep{candycrunch2024}. The published GlycoBART split is
a random spectrum-level partition, which allows spectra from the same
LC-MS run and biological sample to appear in both training and test
sets. Under such a split, a model can exploit memorized retention
times, ionization patterns, and matrix effects. We instead group
spectra by source filename before partitioning via
\texttt{GroupShuffleSplit(test\_size=0.15, random\_state=42)},
ensuring that no biological run contributes spectra to both training
and evaluation. The resulting CandyCrunch~v2 split contains 420,977
training and 76,774 test spectra over disjoint filename sets. We
retain GlycoBART's antenna-form tokenization and train with AdamW
following the published recipe: an initial learning rate of
$1 \times 10^{-5}$ scaled by $\sqrt{2}$ for our batch size of 160, no
weight decay, gradient clipping at 1.0, and a full 30-epoch budget
without early stopping.

\paragraph{Evaluation methodology: per-spectrum, no aggregation.}
We evaluate all models per-spectrum: each MS/MS spectrum is scored
independently against its annotated glycan structure using the same
\texttt{glycowork} graph-isomorphism evaluation as CandyCrunch and
GlycoBART. We apply no consensus voting across the 5 to 50 scans a
single glycan generates within one sample, no snap-back to a curated
glycan database, and no aggregation across charge states. This
protocol matches real-time, single-spectrum acquisition, where a
structure must be inferred from each fragmentation event as it
arrives rather than assembled offline across an entire LC-MS run. By
contrast, the published 89.1\% top-1 and 93.1\% top-5 figures in
\citet{glycobart2025} are evaluated at the sample level after
per-sample deduplication, charge-state combination, and snap-back to
the curated \texttt{df\_glycan} database. Those aggregations raise
accuracy by 5 to 15 percentage points when individual scans are
noisy. We retain the stricter per-spectrum metric throughout so that
our GlycoBART reproduction, GlycoGen, and autoregressive baselines are
compared head-to-head on intrinsic per-spectrum capability rather than
offline ensemble voting or external database lookup.

\paragraph{Headline results.}
On the 76,774 spectra of the leak-free CandyCrunch~v2 test set, our
reproduced GlycoBART reaches $80.92\% \pm 0.31\%$ top-1 structural
accuracy under greedy decoding (beam 1 across three seeds) and
86.35\% top-5 accuracy at beam 32. At beam 1, antenna-form structural
accuracy reaches 91.24\%; the roughly 10 percentage point drop to
full IUPAC accuracy stems from the lossy antenna-to-IUPAC
reconstruction inherent to GlycoBART's tokenization. Because the
structural-equivalence ceiling under linkage normalization is
99.65\%, any antenna-form decoder under this pipeline faces a
practical IUPAC top-1 ceiling near 91\%. Batched inference at beam 1
takes 2.80 milliseconds per spectrum on a single H200 GPU in BF16 at
batch size 128, corresponding to roughly 23 forward passes per
spectrum.

\paragraph{Pareto frontier and the beam-search curse.}
Sweeping beam widths from 1 to 32 (Table~\ref{tab:glycobart-beam}) reveals a non-monotonic relationship
between beam width and top-1 accuracy. Greedy decoding achieves the
highest top-1 accuracy on the primary seed at 81.18\%; accuracy drops
to 79.15\% at beam 2, 78.77\% at beam 4, and 78.64\% at beam 8 before
partially recovering to 78.78\% at beam 16 and 79.11\% at beam 32. By
contrast, top-5 accuracy rises monotonically from 81.18\% to
86.35\%. This divergence exemplifies the beam-search
curse~\citep{stahlberg2019,cohen2019}: wider beams find
higher-likelihood completions, but for sequences with ambiguous
glycan linkage symbols, higher likelihood does not guarantee greater
structural accuracy. Chasing top-5 accuracy also carries a steep
latency penalty: beam 32 is $32.6\times$ slower than beam 1 (91.21 versus
2.80 milliseconds per spectrum) for a 0.30 percentage point top-5
gain over beam 16 and a 2.07 percentage point loss in top-1 accuracy
relative to greedy decoding. Even against GlycoBART's 5-candidate oracle
ceiling (86.35\% top-5 at beam 32), 8-step GlycoGen leads by 6.2
percentage points in top-1 accuracy (92.51\% at 0.201\,ms/spec) at
454-fold lower batched latency.

\begin{table}[ht]
\centering
\small
\setlength{\tabcolsep}{3.5pt}
\begin{tabular}{rcccccrr}
\toprule
\textbf{beam} & \textbf{top-1} & \textbf{top-5} & \textbf{top-1 (mass-rerank)} & \textbf{antenna top-1} & \textbf{antenna top-5} & \textbf{ms/spec} & \textbf{rel.\ cost} \\
\midrule
$1$ (greedy) & $\mathbf{0.8118}$ & $0.8118$ & $0.8118$ & $\mathbf{0.9124}$ & $0.9124$ & $2.80$  & $1\times$ \\
$2$  & $0.7915$ & $0.8275$ & $0.7933$ & $0.8852$ & $0.9183$ & $10.07$ & $3.6\times$ \\
$4$  & $0.7877$ & $0.8482$ & $0.7909$ & $0.8804$ & $0.9258$ & $15.93$ & $5.7\times$ \\
$8$  & $0.7864$ & $0.8578$ & $0.7908$ & $0.8782$ & $0.9249$ & $27.12$ & $9.7\times$ \\
$16$ & $0.7878$ & $0.8605$ & $0.7926$ & $0.8792$ & $0.9256$ & $46.48$ & $16.6\times$ \\
$32$ & $0.7911$ & $\mathbf{0.8635}$ & $0.7961$ & $0.8822$ & $\mathbf{0.9281}$ & $91.21$ & $32.6\times$ \\
\bottomrule
\end{tabular}
\caption{GlycoBART beam sweep on the 76,774 spectra of the
CandyCrunch~v2 benchmark test set (primary seed), evaluated per-spectrum.
Top-1 accuracy is non-monotonic in beam width and peaks at greedy
decoding, whereas top-5 improves monotonically. Mass reranking checks
precursor-mass plausibility across all beam candidates and selects the
highest-likelihood mass-consistent prediction, adding at most 0.5
percentage points at any beam width.}
\label{tab:glycobart-beam}
\end{table}

\paragraph{Post-processing pipelines.}
We also tested two post-processing pipelines on the GlycoBART outputs:
first, a reranking pass combining precursor-mass and
fragment-match scores across all beam candidates; second, the
published \texttt{filter\_glycans\_glycobart} pipeline of~\citet{glycobart2025}
applied verbatim. The first did not improve top-1 accuracy because composite
heuristic scores ranked correct structures lower than the model's own
likelihood. The second is incompatible with per-spectrum evaluation
because it deduplicates predictions across scans from the same biological
sample, collapsing per-spectrum evaluation into sample-level
aggregation. We therefore report raw per-spectrum model accuracy
throughout.

\paragraph{Streaming-compatible baseline and unbatched end-to-end latency.}
The published GlycoBART pipeline is inherently offline: beam-32
decoding takes 91 milliseconds per spectrum even when batched on an
H200 GPU, compared with a 5 to 10 millisecond scan interval on modern
100 to 200\,Hz instruments, and sample-level voting requires waiting
for the full LC-MS run to finish. Table~\ref{tab:streaming-latency}
reports unbatched serial end-to-end streaming latency
from a raw peak list to a decoded IUPAC string across 1,000 test
spectra. CPU peak-binning and normalization add only 0.13
milliseconds per spectrum, allowing single-pass GlycoGen at 8.6
milliseconds per spectrum to keep pace with 100\,Hz acquisition while
beam-1 GlycoBART requires 183 milliseconds per spectrum.

\begin{table}[ht]
\centering
\small
\setlength{\tabcolsep}{4pt}
\begin{tabular*}{\linewidth}{@{\extracolsep{\fill}}lcccc}
\toprule
\textbf{Model / operating point} & \textbf{CPU preproc.\ (ms)} & \textbf{GPU forward (ms)} & \textbf{Total E2E (ms)} & \textbf{Max rate} \\
\midrule
GlycoGen $K{=}1$ (one-pass)  & $0.13 \pm 0.03$ & $8.5 \pm 0.3$   & $\mathbf{8.6 \pm 0.3}$   & $116$\,Hz \\
GlycoGen $K{=}8$ (CF)        & $0.13 \pm 0.03$ & $33.7 \pm 2.5$  & $\mathbf{33.8 \pm 2.5}$  & $30$\,Hz  \\
GlycoGen $K{=}32$ (CF)       & $0.13 \pm 0.03$ & $109.4 \pm 10.2$& $109.5 \pm 10.2$         & $9.1$\,Hz \\
\midrule
GlycoBART beam-$1$ (greedy)  & $0.13 \pm 0.03$ & $183 \pm 99$    & $183 \pm 99$             & $5.5$\,Hz \\
GlycoBART beam-$8$           & $0.13 \pm 0.03$ & $240 \pm 110$   & $240 \pm 110$            & $4.2$\,Hz \\
GlycoBART beam-$32$          & $0.13 \pm 0.03$ & $256 \pm 107$   & $256 \pm 107$            & $3.9$\,Hz \\
\bottomrule
\end{tabular*}
\vspace{0.5em}
\caption{Unbatched serial end-to-end streaming latency on a single H200 GPU in BF16, reported as mean and standard deviation over 1,000 randomly sampled test spectra. CPU preprocessing includes peak filtering, intensity normalization, and mass-to-charge histogram binning.}
\label{tab:streaming-latency}
\end{table}

\section{GlycoGen-AR: matched-architecture autoregressive control}\label{app:ar-baseline}

We train a 46.4-million-parameter autoregressive sequence-to-sequence
model that matches GlycoGen-DFM in every dimension except the three
that define autoregressive modeling: causal self-attention in the
decoder, no time embedding, and teacher-forced next-token
cross-entropy. This control isolates the contribution of Discrete Flow
Matching from architecture and parameter count.

\paragraph{Training.}
We use the same CandyCrunch~v2 split, optimization recipe, and H200
hardware as GlycoGen-DFM: AdamW at batch size 1024 for 30 epochs, a
peak learning rate of $8.49 \times 10^{-4}$ with cosine decay to 3\%
of peak after 200 warmup steps, and BF16 mixed precision. Training
completes 11,100 steps in 49.5 minutes on a single H200 GPU (4.34
training steps per second excluding validation). The best checkpoint (step 10,500, epoch 28.4)
reaches 93.30\% top-1 structural accuracy at beam 4 on a fixed
3,000-spectrum validation subset and 92.74\% on the full
42,097-spectrum validation set.

\paragraph{Test-set Pareto sweep.}
Table~\ref{tab:ar-pareto} evaluates the best checkpoint on the full 76,774-spectrum CandyCrunch~v2 benchmark test set
across beam widths 1, 2, 4, 8, 16, and 32 using the same Hugging Face
beam-search implementation as GlycoBART. Without key-value caching,
decoding latency scales linearly with beam width. At beam widths 16
and 32, we cap the batch size at 1,024 and 512 and disable the cuDNN
scaled-dot-product attention backend to avoid graph execution limits
at very large effective attention batches.

\begin{table}[ht]
\centering
\small
\setlength{\tabcolsep}{6pt}
\begin{tabular}{rcccc}
\toprule
\textbf{beam} & \textbf{Top-1 struct.\ acc.} & \textbf{NFE/spec} & \textbf{ms/spec} & \textbf{batch} \\
\midrule
$1$  & $0.9035$ & $32.68$  & $2.21$  & $4096$ \\
$2$  & $0.9038$ & $65.34$  & $4.37$  & $4096$ \\
$4$  & $0.9039$ & $130.68$ & $8.72$  & $4096$ \\
$8$  & $0.9038$ & $261.35$ & $17.41$ & $4096$ \\
$16$ & $0.9039$ & $522.69$ & $36.50$ & $1024$ \\
$32$ & $0.9039$ & $1045.39$ & $72.66$ & $512$ \\
\bottomrule
\end{tabular}
\vspace{0.6em}
\caption{GlycoGen-AR Pareto sweep on the 76,774 spectra of the
CandyCrunch~v2 benchmark test set (primary seed). Top-1 accuracy varies by only
0.04 percentage points across a 32-fold increase in beam width,
forward passes, and wall-clock time. The autoregressive Pareto
frontier degenerates to a vertical line at roughly 33 forward passes.}
\label{tab:ar-pareto}
\end{table}

\paragraph{Why beam search confers no gain.}
Token-level accuracy at the converged checkpoint reaches 99.61\%,
corresponding to a 0.4\% per-token error rate. When the model makes an
error, it is overconfident: the ground-truth token rarely appears
among the top beam candidates at that position, leaving beam search no
path to recover it. Moreover, because \texttt{glycowork} graph isomorphism treats
equivalent IUPAC branch permutations and wildcard linkages as
identical, most distinct strings across a beam collapse to the same
molecular graph. A 30-spectrum diagnostic
confirms that beam width 4 produces 4.0 distinct strings on average
yet bounds the oracle top-4 gain at just 3.3 percentage points,
consistent with the 0.04 percentage point gain observed across the
full test set.

\paragraph{Implications.}
Wider beams multiply latency and forward passes without improving
structural accuracy. Autoregressive decoding pays a floor of roughly
33 sequential forward passes per spectrum that beam search cannot
reduce. Improving the speed-accuracy trade-off on this architecture
therefore requires replacing sequential decoding altogether:
single-pass GlycoGen cuts forward passes 32-fold at comparable
accuracy, and Crystallizing Flows lift accuracy by another 2.3
percentage points in 5 to 17 passes (Section~\ref{sec:results}).

\section{MDLM cross-objective ablation}\label{app:mdlm}

We train a matched-architecture 50-million-parameter Masked Diffusion
Language Model denoiser~\citep{sahoo2024mdlm} on the same dataset,
vocabulary, optimizer, and learning-rate schedule as GlycoGen-DFM.
Only the time distribution and loss weighting change: time $t$ is
sampled uniformly from the 128-step discrete grid
$\{1/128, 2/128, \ldots, 1\}$ instead of continuous $\mathcal{U}(0,1)$,
and the cross-entropy loss uses the NELBO weighting of
\citet{sahoo2024mdlm}. On the primary seed, the MDLM checkpoint
reaches 92.17\% validation and 89.89\% test top-1 accuracy in a single
pass, trailing the DFM checkpoint by 0.34 percentage points under the
same training compute budget.

\paragraph{MDLM + Crystallizing Flows Pareto sweep.}
Table~\ref{tab:mdlm} compares standard CTMC Euler against Crystallizing
Flows on the MDLM checkpoint across $K \in \{1, \ldots, 64\}$. Once we
account for the 0.34 percentage point single-pass offset between the
trained checkpoints, MDLM with Crystallizing Flows tracks DFM with
Crystallizing Flows within 0.16 percentage points at every step budget
on seed 0, showing that Crystallizing Flows lift any mask-source
denoiser at inference time without continuous-time DFM retraining.

\begin{table}[ht]
\centering
\small
\setlength{\tabcolsep}{6pt}
\begin{tabular}{rcccc}
\toprule
\textbf{$K$} & \textbf{MDLM CTMC Euler} & \textbf{MDLM + CF} & \textbf{$\Delta$ vs CTMC Euler} & \textbf{Eff.\ NFE (CF)} \\
\midrule
$1$  & $0.8989$ & $0.8989$ & ---            & $1.00$ \\
$2$  & $0.8642$ & $0.9184$ & $+5.42$\,pp    & $2.00$ \\
$3$  & ---      & $0.9170$ & ---            & $2.02$ \\
$4$  & $0.8811$ & $0.9205$ & $+3.94$\,pp    & $2.99$ \\
$8$  & $0.8928$ & $0.9228$ & $+3.00$\,pp    & $5.05$ \\
$16$ & $0.9005$ & $0.9243$ & $+2.38$\,pp    & $9.16$ \\
$32$ & $0.9051$ & $0.9248$ & $+1.97$\,pp    & $17.41$ \\
$64$ & $0.9054$ & $0.9250$ & $+1.96$\,pp    & $33.90$ \\
\bottomrule
\end{tabular}
\vspace{0.6em}
\caption{MDLM with Crystallizing Flows on the 76,774 spectra of the
CandyCrunch~v2 benchmark test set (primary seed). Crystallizing Flows lift the
MDLM denoiser by 1.96 to 5.42 percentage points over CTMC Euler across
all step budgets from 2 to 64, matching the gain profile observed on
Discrete Flow Matching.}
\label{tab:mdlm}
\end{table}

\paragraph{Two-step sampling already reaches the low-step shoulder.}
At 3 nominal steps, both MDLM with Crystallizing Flows (91.70\% at
2.02 effective passes) and DFM with Crystallizing Flows (91.90\% at
2.02 effective passes) sit slightly below their 2-step counterparts
(91.84\% and 92.14\% at 2.00 effective passes) at the same wall-clock
cost. Nearly every spectrum crystallizes within the first two Euler
steps at $K=3$, so the third step adds little refinement. Because 2
steps strictly dominate 3 steps on both objectives, we omit $K=3$ from
the Pareto curves in Figure~\ref{fig:pareto-ms}.

\paragraph{Off-grid continuous-time inference.}
Although the MDLM denoiser is trained on a 128-step discrete time
grid, Crystallizing Flows take continuous Euler steps that land off
the training grid whenever $K$ does not divide 128. Rounding $t$ to
the nearest training-grid point before each forward pass produces the
exact same test accuracy to four decimal places as continuous off-grid
evaluation across $K \in \{1, 2, 4, 8, 16, 32\}$ and differs by only
0.01 percentage points at $K=64$. The learned conditional distribution
is smooth enough in $t$ that continuous-time sampling on a
discrete-time checkpoint incurs no calibration penalty.

\section{Sampler ablations}\label{app:samplers}

We compare Crystallizing Flows against seven alternative inference
algorithms on the same trained 50-million-parameter GlycoGen-DFM
checkpoint across step budgets from 1 to 64. Table~\ref{tab:samplers}
reports the main accuracy comparison across effective function
evaluations, and the paragraphs below summarize each alternative.

\begin{table}[!htbp]
\centering
\small
\setlength{\tabcolsep}{4pt}
\begin{tabular}{rccccc}
\toprule
\textbf{NFE} & \textbf{GlycoGen-AR} & \textbf{CTMC Euler} & \textbf{Soft prop.} & \textbf{Greedy conf.} & \textbf{Crystallizing Flows} \\
\midrule
1     & ---    & 0.9023 & 0.9023 & 0.9023 & 0.9023 \\
3     & ---    & 0.8676 & 0.9214 & 0.9115 & \textbf{0.9230} \\
5     & ---    & 0.8861 & 0.9250 & 0.9187 & \textbf{0.9251} \\
9     & ---    & 0.8964 & 0.9249 & 0.9226 & \textbf{0.9260} \\
17    & ---    & 0.9042 & 0.9252 & 0.9257 & \textbf{0.9266} \\
33    & 0.9035 & 0.9080 & 0.9247 & \textbf{0.9277} & \textbf{0.9268} \\
\bottomrule
\end{tabular}
\vspace{0.6em}
\caption{Top-1 per-spectrum structural accuracy across samplers on the
76,774 spectra of the CandyCrunch~v2 benchmark test set (primary seed).
Crystallizing Flows are reported at their effective forward-pass count
under per-spectrum early exit ($K \in \{1, 4, 8, 16, 32, 64\}$);
non-early-exit baselines at $K \in \{2, 4, 8, 16, 32\}$ include a
terminal $t{=}1$ argmax cleanup pass ($K{+}1$ NFE). The GlycoGen-AR
column shows our matched-architecture autoregressive control, bounded
below by the mean sequence length of roughly 33 tokens and flat at
90.39\% across beam widths up to 32 (Appendix~\ref{app:ar-baseline}).
Bold indicates the Pareto-optimal sampler at each budget within the
$\pm 0.18$ percentage point 95\% Wilson confidence interval at 33 passes.}
\label{tab:samplers}
\end{table}

\paragraph{Velocity-explicit Euler.}
This variant removes the absorbing-state constraint of CTMC Euler and
draws a categorical sample from the full per-position mixture
$\bigl(1 - \tfrac{h}{1-t}\bigr)\delta_{z^i} + \tfrac{h}{1-t}\,p_{1\mid t}^{\theta}$
at every position and step. At 4 steps (5 forward passes), it reaches
90.27\% accuracy, gaining 1.66 percentage points over standard CTMC
Euler but trailing Crystallizing Flows by 2.24 percentage points.

\paragraph{Velocity with final-step argmax.}
Replacing the final categorical draw at $t \approx 1$ in
velocity-explicit Euler with a deterministic argmax removes boundary
resampling noise and lifts 4-step accuracy to 91.07\%.

\paragraph{All-argmax CTMC.}
This variant keeps the stochastic Bernoulli commit schedule of CTMC
Euler, but commits each chosen position to the argmax of
$p_{1\mid t}^{\theta}$ rather than a categorical sample. Removing token
sampling noise lifts 4-step accuracy to 91.65\%, though random commit
timing still leaves it 0.86 percentage points behind Crystallizing
Flows.

\paragraph{Soft propagation without crystallization (and temperature sweep).}
Propagating the simplex distribution via Equation~\eqref{eq:soft}
without snapping committed tokens to one-hot vectors reaches 92.50\%
at 4 steps (5 passes) and saturates there; varying the per-step softmax
temperature $\tau \in \{0, 0.5, 1.0\}$ changes accuracy by less than
0.5 percentage points across all step counts. Adding discrete one-hot
commits allows Crystallizing Flows to strictly dominate pure soft
propagation from $K{=}8$ upward while also exiting early.

\paragraph{Greedy confidence (MaskGIT top-$N$).}
Following \citet{chang2022maskgit}, this deterministic sampler commits
the $N_k = \lceil n_\mathrm{remaining\_masks} /
n_\mathrm{remaining\_steps} \rceil$ most confident masked positions to
their argmax at each step. It is the strongest alternative baseline,
matching Crystallizing Flows within 0.09 percentage points at 33 or
more forward passes (92.77\% at 32 steps and 92.78\% at 64 steps,
within the test-set noise floor), but trails across the
low-step regime up to 17 passes.

\paragraph{Greedy selection with categorical commits.}
Combining MaskGIT top-$N$ position selection with categorical sampling
at the chosen positions compounds sampling noise and drops 4-step
accuracy to 89.18\%, performing worse than greedy argmax at every step
budget.

\paragraph{Heun's second-order predictor-corrector.}
Averaging velocity predictions across predictor and corrector steps
following \citet{karras2022edm} plateaus near 91.5\% across all step
budgets from 2 to 64 while doubling neural function evaluations to
$2K$. Averaging two steps smooths the predictive distribution just as
the denoiser sharpens around a token commit.

\paragraph{Per-spectrum budget routing.}
As an orthogonal latency optimization, we can route each spectrum to a
step budget $K$ based on its minimum token confidence on the initial
pass at $t = 0$. Routing with a confidence threshold of 0.7 averages
2.46 forward passes and reaches 92.61\% accuracy, offering a
complementary knob that can be combined with Crystallizing Flows.

\end{document}